\documentclass[preprint]{aastex}	

\newcommand{\czhel}{\mbox{$cz_{\rm hel}$}}	
\newcommand{\etal}{et al.}			
\newcommand{\hbgdav}{\mbox{$\langle{\rm H}\beta\gamma\delta\rangle$}}
\newcommand{\hst}{{\it HST\,}}			
\newcommand{\kms}{km~s$^{-1}$}			
\newcommand{\kratio}{K/(H$\epsilon$\,+\,H8)}	
\newcommand{\mass}{\mbox{${\cal M}$}}		
\newcommand{\msun}{\mbox{${\cal M}_{\odot}$}}	
\newcommand{\mgb}{Mg~$b$}			
\newcommand{\n}{NGC~}				
\newcommand{\reff}{\mbox{$R_{\rm eff}$}}	
\newcommand{\tcross}{\mbox{$t_{\rm cr}$}}	
\newcommand{\vizero}{\mbox{$(V\!-\!I)_0$}}	
\newcommand{\zsun}{\mbox{$Z_{\odot}$}}		

\shorttitle{Young Globular Clusters in NGC 3921}
\shortauthors{Schweizer, Seitzer, \& Brodie}

\begin{document}

\title{KECK SPECTROSCOPY OF TWO YOUNG GLOBULAR
       CLUSTERS IN THE MERGER REMNANT NGC 3921}

\author{Fran\c cois Schweizer\altaffilmark{1}}
\affil{Carnegie Observatories, 813 Santa Barbara Street, Pasadena, CA
91101; schweizer@ociw.edu}

\author{Patrick Seitzer\altaffilmark{1}}
\affil{Department of Astronomy, University of Michigan, 818 Dennison Building,\\
Ann Arbor, MI 48109; pseitzer@umich.edu}

\and

\author{Jean P. Brodie}
\affil{UCO/Lick Observatory, University of California, Santa Cruz, CA 95064;
brodie@ucolick.org}

%
\altaffiltext{1}{Guest Observer, Keck Observatory.}


\begin{abstract}
\noindent
Low-resolution, ultraviolet-to-visual spectra of two candidate globular
clusters in the merger remnant \n3921 ($cz_{_{\rm LG}}=6021$ \kms) are
presented.  These two clusters of apparent magnitude $V\approx 22.2$
($M_V\approx -12.5$) lie at
projected distances of about 5~kpc (0.9\reff) from the center and move
with halo-type radial velocities relative to the local galaxy background.
Their spectra show strong Balmer absorption lines
[EW(H$\beta$\,--\,H$\delta$)~= 11\,--\,13 \AA] indicative of
main-sequence turnoffs dominated by A-type stars.  Comparisons with
model-cluster spectra computed by Bruzual \& Charlot and others
yield cluster ages in the range of 200\,--\,530~Myr, and metallicities
about solar to within a factor of three.  Given their small half-light
radii ($\reff\la 5$~pc) and ages corresponding to $\sim$10$^2$ core-crossing
times, these clusters are gravitationally bound and, hence, indeed young
globulars. Assuming that they had Chabrier-type initial mass
functions, their estimated current masses are 
$(2.3\pm 0.1)\times 10^6 \msun$ and $(1.5\pm 0.1)\times 10^6 \msun$,
respectively, or roughly half the mass of $\omega$ Cen.
Since \n3921 itself shows many signs of being a $0.7\pm 0.3$~Gyr
old protoelliptical, these two young globulars of roughly solar metallicity
and their many counterparts observed with the {\it Hubble Space Telescope}
provide supporting evidence that, in the process of forming elliptical-like
remnants, major mergers of gas-rich disks can also increase the number
of metal-rich globular clusters.

\end{abstract}



\section{INTRODUCTION}

During the past decade it has become increasingly apparent that
galactic mergers, bulge building, and the formation of large numbers of
star clusters are closely intertwined subjects.  Evidence for systems of
luminous blue clusters newly formed in mergers has accumulated both before
and especially since the launch of the {\it Hubble Space Telescope (HST)}.
It now appears that of the thousands of clusters formed in ongoing mergers
of gas-rich galaxies (e.g., \n4038/39: \citealt{ws95}, \citealt{whit99};
\n3256: \citealt{zepf99}) only the densest and more massive will survive
for $\ga$500 Myr \citep{fall01} and become the young globular clusters
seen in merger remnants of recent vintage.

Observations of globular cluster (GC) systems in recent remnants such as
\n1275 \citep{holt92,carl98},
\n3597 \citep{holt96,carl99},
\n3921 (\citealt{schw96b}, hereafter SMWF96), and
\n7252 \citep{whit93,mill97}
are especially valuable because in these remnants the starbursts have
subsided, the dense gas and dust obscuring the view have diminished,
and most of the freshly minted clusters have evolved sufficiently to
reveal their true nature.  Over the several 100 Myr a major disk--disk
merger takes to complete, loose star clusters and associations tend to
disperse, while gravitationally-bound dense clusters survive as a fossil
record of the merger's star-formation history.  Hence, determining the
ages and metallicities of these surviving clusters is an important
first step in trying to reconstruct that history.

This goal has taken added importance since the discovery of bimodal
color and metallicity distributions in the GC systems of many giant
ellipticals, predicted by \citet{ashm92} from merger models of
E formation. Since in such ellipticals the red globulars typically have
metallicities of $Z\approx 0.2$\,--\,2 \zsun\ (e.g., \citealt{forb01};
\citealt{peng04}), a question of interest is just how metal-rich the
second-generation GCs formed during major mergers are.  There is growing
evidence that such clusters have roughly solar metallicities, as measured
to date for two young globulars in \n7252 \citep{ss93,ss98}, one such
cluster in \n1275 \citep{zepf95,brod98}, three intermediate-age globulars
in \n1316 \citep{goud01a}, and two in \n3610 \citep{stra03,stra04}.

The present paper describes exploratory spectroscopic observations aimed
at verifying the nature of two candidate young globular clusters in
\n3921, and at determining their ages and metallicities.  Similar to, but
slightly younger than the more well-known \n7252, \n3921 is a very recent
remnant of two merged disk galaxies \citep{tt72} and shows many signs
of being a $0.7\pm 0.3$ Gyr old protoelliptical (\citealt{schw96a},
hereafter S96).  Predicted to fade by only $\sim$1.5 mag from
its present $M_V = -22.1$ over the next 12 Gyr, this remnant is likely
to remain a luminous field elliptical.  Its young candidate GCs and
looser associations were first discovered from the ground (S96)
and then in larger numbers with \hst\ (SMWF96).  Our spectroscopic
observations pertain to two of its brightest candidate globulars.

\n3921 itself is located at $\alpha_{\rm J2000}=11^{\rm h}51^{\rm m}07\fs0$,
$\delta_{{\rm J}2000}=+55\degr04\arcmin43\arcsec$ (SMWF96) and has a
recession velocity relative to the Local Group of
$cz_{_{\rm LG}} = +6021\pm 15$ \kms\
(S96), which places it at a distance of 86 Mpc for $H_0 = 70$
\kms\ Mpc$^{-1}$.  At that distance, adopted throughout the present
paper, $1\arcsec = 417$ pc.  The corresponding
distance modulus is $(m-M)_0 = 34.67$. The Milky Way foreground extinction
is small, with values in the literature ranging between $A_V=0.00$
(\citealt{rc3}) and 0.047 \citep{schl98}.  We adopt the mean value
of $A_V = 0.023$, with which the absolute visual magnitude of \n3921
becomes $M_V = -22.09$.

In the following, \S 2 describes the observations and reductions,
\S 3 presents results concerning the kinematics, ages, and mean
metallicity of the two young clusters, and \S 4 discusses issues
concerning the clusters' physical nature and origin.  Finally,
\S 5 summarizes our main conclusions.

\section{OBSERVATIONS AND REDUCTIONS}

Two candidate GCs in \n3921 were observed with the Low-Resolution Imaging
Spectrograph (LRIS, \citealt{oke95}) of the Keck~I telescope on Mauna Kea
on 2000 February 29.  During a three-hour window of opportunity
beginning shortly after midnight, we realigned the mirror optics
of the telescope and placed the long slit of LRIS ($165\arcsec\times
1\farcs0$) at position angle P.A.\ = $29\fdg69$ across the two candidate
clusters via measured offsets from the nucleus.  Figure 1 identifies the
clusters, S1 and S2, and shows the position of the LRIS slit across them.

Table 1 presents basic data about these two clusters, including their
offsets and distances from the nucleus, effective radii, magnitudes and
colors, and the radial velocities measured in the present study.  Clusters
S1 and S2 were the two brightest candidate globulars discovered from
ground-based observations (S96) and are the third and fourth brightest
candidate GCs cataloged from observations obtained with {\it HST\,} and
WFPC2 (SMWF96).

Figure 2 shows the locations of S1 and S2 in a color--magnitude diagram of
the 102 candidate GCs found with \hst.  The two objects lie right on the
`main' sequence of candidate young clusters and are, with absolute magnitudes
of $M_V \approx -12.5$, very luminous.  Of the two still brighter objects,
one (\#54 in SMWF96) lies only $2\farcs7$ from the exceedingly bright
nucleus of the galaxy, too close for spectroscopic observation with LRIS.
The other (\#2 in SMWF96) lies at $50\farcs2$ (corresponding to
$r_{\rm proj} = 21$ kpc) from the nucleus and may, despite its
young-cluster-like color, be a foreground star (S96).

After aligning the spectrograph slit across clusters S1 and S2, we obtained
four consecutive 30-min exposures of these objects, followed by a short
calibration exposure with a Ne-Kr-Xe-Hg emission-line source. The exposures
were centered approximately on meridian transit, with the airmass varying
only between 1.23 and 1.26.  The seeing remained in the range
$0\farcs7$\,--\,$0\farcs9$ all night.  Two standard stars plus several
velocity standards were observed as well.  During our observations, LRIS
was equipped with a 600 g mm$^{-1}$ grating blazed at 5000 \AA\ and a
Tektronix 2K$\,\times\,$2K chip as detector.  This combination provided
a mean reciprocal dispersion of 1.24 \AA\ pixel$^{-1}$, a spectral
resolution of about 5.5 \AA, and a wavelength coverage of 3750\,--\,6280 \AA.
The scale perpendicular to the dispersion was $0\farcs213$ pixel$^{-1}$.

All reductions were performed with IRAF\footnote{
The Image Reduction and Analysis Facility (IRAF) is distributed by the
National Optical Astronomy Observatories (NOAO), which are operated by
the Association of Universities for Research in Astronomy (AURA), Inc.,
under a cooperative agreement with the National Science Foundation.}
and its spectral-extraction tasks.  First the data frames were debiased,
flat-fielded, cleaned of cosmic-ray defects, wavelength calibrated, and
rectified.  Then the spectra of the two clusters were traced on each
frame and extracted in a $1\farcs28$ (= 6 pixel) wide band with simultaneous
subtraction of sky and galaxy background.  Because the cluster counts
contribute only 5\,--\,10\% of the total signal at each wavelength and the
galaxy background is uneven, the background subtraction had to be done
with extreme care.

For Cluster S1, a judicious choice of background-extraction bands on either
side allowed us to use parabolic fits to the galaxy-plus-sky background at
each wavelength.  Cluster S2, however, sits precariously on an edge
of the galaxy light distribution, where the gradient changes abruptly.
Again using background-extraction bands on either side of the cluster, we
had to use 4th-order polynomials to achieve proper background subtraction.
The background fits were monitored by visual inspection at several
wavelengths, and linear background fits with two narrow background strips
closely flanking the cluster were performed as a check.  The cluster spectra
extracted with linear background subtraction closely resemble those with
4th-order background subtraction, but are more noisy because of the poorer
background statistics.  Hence, only the spectra with 4th-order background
subtraction were used for the subsequent analysis.

As a second check on the success of our background subtractions we
also extracted a $4\farcs9$ (= 23 pixel) wide strip of galaxy background
lying midway between the two clusters, at $\sim$4 kpc ($9\farcs5$) from
the nucleus.  The night-sky portion of the background spectrum was removed
by subtracting scaled night-sky spectra extracted near both ends of the long
slit. As we show below, the spectrum of the galaxy background differs
strongly from the two cluster spectra and yields a significantly lower
radial velocity, indicating that contamination of the cluster
spectra by galaxy background---if present---is insignificant.

For each cluster and the galaxy background we produced a final spectrum
by coadding the spectra extracted from three of the four half-hour exposures. 
The omitted frame, corresponding to the first exposure, suffered from poor
image quality, likely due to intermittent mirror-alignment problems,
and would have contributed more noise than signal to the final cluster
spectra. Before coadding them,  the individual spectra were corrected for
minor shifts ($\la 0.5$ \AA) due to flexure, as determined from night-sky
emission lines.  Finally, the sum spectra were corrected for atmospheric
extinction at Mauna Kea and flux calibrated via the two observed standard
stars, PG 0823+546 and PG 0939+262 \citep{mass88}.

Figure 3 displays the flux-calibrated spectra of S1, S2, and the galaxy
background in between, plotted versus rest wavelength.  In the spectra
of the two clusters, note the strong Balmer absorption lines indicative
of A-type main-sequence stars; on expanded plots, these lines are visible
up to H13. Except for the \ion{Ca}{2} K-line, the metal lines appear weak.
In contrast, the spectrum of the galaxy background shows weaker Balmer
lines, much stronger metal lines, and a continuum indicative of more
composite and older stellar populations.

\section{RESULTS}

\subsection{Cluster Velocities}

Table 1 gives the heliocentric radial velocities \czhel\ measured for
clusters S1 and S2.  These velocities represent averages of velocities
measured by two different methods.  First, we determined a mean velocity
from five Balmer lines (H$\beta$--H$\epsilon$, H9) measured individually
in each cluster spectrum. And second, we determined a mean velocity for
each cluster via cross-correlation.  Since none of the velocity standards
observed during the run were of sufficiently early spectral type to
yield a good match to the spectra of S1 and S2, we used as a template the
spectrum of the cluster NGC\,7252:\,W3 obtained years earlier with
the Blanco 4-m telescope at similar resolution \citep{ss98}.
This cluster has a very similar age and spectrum, and its radial
velocity is known with high accuracy, \czhel~= $4822.5 \pm 1.0$ \kms\
(\citealt{mara04}; cf.\ with $4821 \pm 7$ \kms\ by \citealt{ss98}).
Results from the two methods agreed within the combined
errors, and the velocities in Table 1 are the straight averages.

Also given in Table 1 are the cluster velocities $\Delta v$ relative to
the nucleus of \n3921.  These relative line-of-sight velocities were
computed from
$$\Delta v = (cz_{\rm hel} - cz_{\rm hel,sys})/(1+z_{\rm hel,sys}),$$
where the denominator is a relativistic correction and the systemic
velocity of \n3921 is $cz_{\rm hel,sys} = 5926 \pm 15$ \kms\ (S96). The
values of $\Delta v = +114 \pm 34$ \kms\ for S1 and $+51 \pm 25$ \kms\
for S2 clearly indicate that these clusters are physically associated
with \n3921.

For comparison, the patch of galaxy background extracted between the
two clusters (Fig.\ 3) has a line-of-sight velocity of
$\Delta v = -25 \pm 26$ \kms\ relative to the nucleus.  Given the
relatively large velocity differences between the two clusters and the
local galaxy background, the clusters are likely to belong to a halo
population and show halo kinematics.  Observations of at least six to
eight more clusters are needed to test this hypothesis and determine a
reliable halo velocity dispersion.

\subsection{Cluster Ages from Balmer and H\,+\,K Lines}

The spectra of clusters S1 and S2 have relatively low signal-to-noise
ratios (Table 1), making it difficult to measure metal lines with the
accuracy required for the determination of metallicities.  Fortunately,
the Balmer absorption lines, which contain most of the age information,
are strong and easily measurable.
Therefore, we follow the same two-step procedure as used by \citet{ss98}.
First, we assume that the clusters have approximately
solar metallicities and determine their ages from the Balmer and
H\,+\,K lines via comparison with high-resolution model spectra
for $Z$~= \zsun.  Then, in \S 3.3 we estimate metallicities and ages 
via Lick indices and check that---indeed---the assumption of solar
metallicity appears to be good to within a factor of about three.

To determine the cluster ages, we measured equivalent widths of Balmer
and other strong absorption lines from the observed spectra and compared
them directly with equivalent widths measured in the same manner from
high-resolution model spectra by \citeauthor{bc96} (1996, hereafter BC96;
see also \citealt{bc93,bc03}).  Table~2 gives equivalent widths of
the Balmer and \ion{Ca}{2} H\,+\,K lines measured from the observed
spectra.  The
adopted passbands were 62 \AA\ wide for H$\beta$, 55 \AA\ for H$\gamma$,
52 \AA\ for H$\delta$, 48 \AA\ for H\,+\,H$\epsilon$, 17 \AA\ for K, and
40 \AA\ for H8.  Continuum passbands were chosen on either side of the
line features, and the measurements were carried out with the task
SPINDEX of the software package VISTA \citep{gonz93}. In addition to
the equivalent widths of the measured six features, Table~2 also gives
the mean equivalent width of H$\beta$\,--\,H$\delta$
\newline\medskip
\centerline{\hbgdav~$\equiv$ 
\case{1}{3}[EW(H$\beta$)\,+EW(H$\gamma$)\,+\,EW(H$\delta$)]\,,}
\newline
the line ratio
\newline
\centerline{\kratio~$\equiv$ EW(K)/[EW(H\,+\,H$\epsilon$)\,+\,EW(H8)]\,,}
\newline\noindent
and the derived logarithmic cluster ages.

Figure 4 illustrates the derivation of cluster ages from the measured
equivalent widths and line ratios.  The plotted curves represent the
evolution of \kratio, EW(H$\beta$), and \hbgdav\ as functions of
logarithmic age, as measured from the BC96 high-resolution model-cluster
spectra of solar metallicity (named ``gsHR'').  To compare the
observations with the models, the figure also shows the line ratios and
equivalent widths measured for clusters S1 and S2 as horizontal lines.
The ages of S1 and S2 were then determined from this figure as follows.
Logarithmic ages were read off at the intersections between the horizontal
lines marking the measurements and the curves representing the model line
ratio or equivalent widths. Since in three out of four cases the Balmer
equivalent widths yield two possible values for the age, the ratio
\kratio\ was used to select the more likely of the two values.  A
weighted mean was then formed of the logarithmic ages obtained from
\kratio, EW(H$\beta$), and \hbgdav.  This weighted mean
is the logarithmic cluster age listed in the last column of Table~2.
From these logarithmic ages follows that, for an assumed solar
metallicity, the linear ages of clusters S1 and S2 are
$450^{+80}_{-110}$ Myr and $280^{+70}_{-80}$ Myr, respectively.

\subsection{Cluster Metallicities}

To estimate the metallicities of the two clusters, we measured Lick
line-strength indices \citep{fabe85,gonz93,trag98} from appropriately
smoothed versions of the observed cluster spectra and compared them to
indices computed for model clusters by Bressan, Chiosi, \& Tantalo (1996)
and \citeauthor{bc03} (2003, hereafter BC03).  Specifically, we used
the H$\beta$\,--\,[MgFe] diagram introduced by \citet{gonz93} to estimate
cluster ages and metallicities simultaneously from the measured indices.
The metallicity index [MgFe] is defined through
[MgFe]~$\equiv [{\rm Mg} b\times \case{1}{2} ({\rm Fe5270} +
{\rm Fe5335})]^{1/2}$,
where \mgb, Fe5270, and Fe5335 are Lick indices expressed in angstroms.
This index is nearly insensitive to variations in $\alpha$/Fe,
the ratio of $\alpha$-elements to iron \citep{tmb03}.

Figure 5 shows three versions of the H$\beta$\,--\,[MgFe] diagram, with
isochrones and isometallicity lines based on \citet{bres96} and
BC03 models.
Data points in each diagram mark the positions of S1, S2, a fictitious
cluster with the mean spectrum of S1 and S2, and two bright young
globulars in \n7252 for comparison.  Taken at face value, the positions
of clusters S1 and S2 in these diagrams might seem to suggest that S1
is both younger and more metal poor than S2, thus contradicting our
earlier result---based on assumed solar metallicity---that S1 probably
is the older cluster (\S 3.2).  However, notice that the error bars are
large and that both clusters appear {\it outside} the area covered by
the isochrone--isometallicity grids.  Cluster S2 would have to have a true
metallicity of $Z\ga 3 \zsun$ to fit into extended such grids, while
Cluster S1 lies outside the area covered by any grids, even those covering
younger ages.  Hence, it appears that the errors in the measured Lick
indices of S1 and S2 are too large to draw any firm conclusions about age
and metallicity differences.

For this reason, and to improve the available signal-to-noise ratio,
we averaged the spectra of clusters S1 and S2, measured Lick indices
from the mean spectrum, and plotted the resulting position of a
fictitious mean cluster $\langle$S1+S2$\rangle$ on the
H$\beta$\,--\,[MgFe] diagrams of Fig.~5.  This position suggests,
with still uncomfortably large error bars, that on average the two
clusters are about $10^{8.4\pm 0.2}$ yr = $250^{+150}_{-90}$ Myr old
and have solar to twice solar metallicity (depending on which model
grid one uses) to within a factor of about 2\,--\,3.

Within the large error bars, these independent estimates of age and
metallicity based on the H$\beta$\,--\,[MgFe] diagram agree with the
cluster ages estimated in \S 3.2 and with our former assumption that the
clusters have roughly solar metallicities.  However, given the large
errors especially in metallicity, we regard the ages derived in \S 3.2
and listed in Table~2 as more reliable than those derived from the
H$\beta$\,--\,[MgFe] diagram.  The former ages depend on three
Balmer-line equivalent widths plus the ratio \kratio, while the latter
ages depend mainly on the one Lick H$\beta$ index and, in the first
two of the three H$\beta$\,--\,[MgFe] diagrams, on fitting functions
{\it extrapolated} to ages younger than 1 Gyr.

We conclude that the metallicity is $[Z]\equiv \log(Z/\zsun) =
0.0\pm 0.5$ for both clusters, and that new, significantly higher-$S/N$
spectra with Keck or larger telescopes will be needed to determine
more precise metallicities and possible metallicity differences between
young globular clusters in \n3921.

\section{DISCUSSION}

The present discussion addresses some issues concerning the ages,
metallicity, nature, and origin of the two young clusters in \n3921.
The single most interesting question is whether these clusters truly
are young globulars.

\subsection{Cluster Ages}

Given the ages derived in \S 3.2 and their error bars, it seems likely
that clusters S1 and S2 both have ages falling in the range 200\,--\,530
Myr.  Over the next 13 Gyr, solar-metallicity clusters in this age
range will fade by $\Delta M_V \approx 3.9$\,--\,3.2 mag from stellar
evolution alone (BC03, for \citealt{chab03} initial mass function), and
possibly more if loss of stars due to evaporation is significant
(Fall \& Zhang 2001).  Hence, when S1 and S2 will reach an age comparable
to that of most present-day old globulars, they should have absolute
magnitudes well within the range typical for such globulars.  Specifically,
at age 13.5 Gyr and if evaporation is negligible, Cluster S1 will shine
with $M_V\approx -8.6$ and S2 with $-9.2$.

How secure are the spectroscopically determined cluster ages, and
could---by any chance---the real ages lie well outside the above quoted
range of 200\,--\,530 Myr?  To address these questions, we have performed
three checks.

First, a visual comparison of the cluster spectra shown in Fig.~3 with
sequences of Magellanic-Cloud cluster spectra arranged by age
\citep{bica86,leon00,leon03} suggests immediately that clusters S1 and
S2 do, indeed, have ages within about the quoted range.  The spectra of
younger Magellanic-Cloud clusters ($\tau\approx 10\,$--\,100 Myr) differ
significantly from those of S1 and S2 by having much stronger UV continua
and weaker K lines, while those of clusters older than $\sim$600 Myr have
stronger K lines, weaker Balmer lines, and redder continua.

Second, a more detailed comparison with an age sequence of model-cluster
spectra of solar metallicity (BC03) strengthens this conclusion.
Figure~6 shows the observed spectra of S1 and S2 with BC03 model spectra
superposed.  In each case, we have simply selected from the age sequence
the model spectrum nearest in time to the cluster age and overplotted it.
Even without interpolating the model spectra in time or smoothing them, the
matches between observed and model spectra are impressive.  These matches
deteriorate rapidly when one superposes spectra of model clusters half or
twice as old as those shown.  In addition, we have specifically checked
that during the much earlier, red-supergiant phase of cluster evolution
($\tau \approx 8\,$--\,20 Myr) the model spectra remain significantly
different from the observed ones, with both the Balmer and \ion{Ca}{2}
K lines weaker and He lines showing up that are not seen in S1 or S2.

Hence, these two spectroscopic checks appear to confirm that the true
ages of S1 and S2 do, most likely, lie within the quoted range of
200\,--\,530 Myr.

Yet, a third, photometric check is less successful.  From the cluster
colors \vizero\ given in Table~1 and the same BC03 (or BC96) models
used for the spectroscopic analysis above, we derive photometric
ages about 1.5\,--\,2 times as large as the spectroscopic ages based
on Balmer-line equivalent widths: $\tau_{\rm phot} = 725\pm 48$ Myr
for Cluster S1, and $510\pm 28$ Myr for S2 (BC03 models, $Z=\zsun$,
Chabrier IMF).  At least four different factors could cause this
discrepancy: reddening, non-solar metallicity, an abnormal initial
mass function (IMF), and problems with the models.

We briefly consider these four factors in turn:
(1) To bring the photometric ages into agreement with the spectroscopic
ages, reddening within \n3921 would have to be significant,
$E_{V\!-\!I}=0.15$ for S1 and 0.09 for S2 (corresponding to $A_V=0.36$
and 0.21, respectively), which seems possible given the apparent scatter
in cluster colors seen in Fig.~2.
(2) To achieve the same agreement through non-solar metallicity, the
clusters would have to have $Z\approx 3\, \zsun$, an unusually
high value, but still on the edge of the range permitted by our
observations (\S 3.3).
(3) A top-heavy IMF might yield abnormally strong Balmer absorption
lines, as suspected for young clusters in \n1275 \citep{brod98}, thus
falsifying the spectroscopic ages.
And (4), there is always the possibility that the models used do not
reproduce the cluster colors well during the Asymptotic-Giant-Branch
(AGB) phase of evolution ($\tau=10^8$\,--\,$10^9$ yr, see
\citealt{mara01}), when for clusters of mass $\la${}$10^6 \msun$
stochastic noise in the colors is significant (BC03, esp.\ Fig.~8).
Although we cannot adjudicate between these various possibilities
at present, we suspect that a combination of reddening and stochastic
color noise may be responsible for most of the discrepancy between
the photometric and spectroscopic ages of S1 and S2.

How significant is the age difference between the two clusters, and could
they both have the same age?  The formal age difference derived from the
cluster ages given in \S 3.2 is $\Delta{\rm Age} = 170\pm 130$~Myr, which
implies that this difference is significant only at the 1.3$\,\sigma$ level.
Hence, equal or very similar ages for clusters S1 and S2 cannot be excluded
with any certainty.  Yet, there is no reason to prefer a near-zero
age difference over the most likely 170~Myr age difference, since in
major disk mergers the period of enhanced star and cluster formation lasts
typically about 200\,--\,500~Myr (e.g., \citealt{miho93}, esp.\ Figs.\ 10,
13, and 24; \citealt{barn04}, esp.\ Fig.\ 5).  Hence, even with an age
difference of $170\pm 130$ Myr clusters S1 and S2 are likely to have
formed during the same starburst period of the merger.

How do the newly determined spectroscopic ages of S1 and S2 compare
with what is known about the merger history of \n3921? This remnant of
a likely S0--Sc or Sa--Sc merger \citep{hivg96} is a $0.7\pm 0.3$ Gyr
old protoelliptical (S96).  Its ``E\,+\,A'' spectrum near the center
suggests that a major starburst occurred at some time during a period
$\sim\,$0.5\,--\,1.0 Gyr ago, whereas the central morphological
disequilibrium hints at a somewhat more recent coalescence.  Taken
literally,
the above starburst period and the spectroscopic ages of S1 and S2
would imply that the two clusters formed near the end or even in the
aftermath of the main starburst.  However, the uncertainties are large
enough
that S1 and S2 may well have formed {\it during} the main starburst.
What seems certain is that S1 and S2 are unlikely to have formed
before the main starburst, in agreement with the evidence from
102 candidate GCs observed with \hst\ (SMWF96).

\subsection{Metallicity}

The metallicities of clusters S1 and S2 are about solar to within a
factor of three (\S 3.3).  This is not unexpected, given that the
clusters formed from molecular gas within the past $\sim$0.5 Gyr and
the two galaxies that merged in \n3921 were major disks.

The clusters' near-solar metallicities are also consistent with evidence
that (1) \n3921 itself is a protoelliptical (\citealt{tt72}; S96)
and (2) giant ellipticals tend to have globular-cluster systems with
bimodal metallicity distributions, one peak of which contains ``metal-rich''
clusters of $[Fe/H]\ga -1.0$ (e.g., \citealt{az98,kund01,lars01}).
Observations
with \hst\ indicate that the recent merger in \n3921 produced $\ga$100
new globular clusters, increasing the total number of GCs by at least
40\% (SMWF96). If most of these young GCs are as metal-rich as S1 and S2
are, then the ratio of metal-rich to metal-poor GCs must be $\ga$0.4,
in rough agreement with the average ratio and 1$\sigma$-range of
$0.8\pm 0.4$ observed in normal elliptical galaxies \citep{forb97,lars01}.
Therefore, within a few gigayears \n3921 may possess a GC system not
unlike that of a giant field elliptical. The metallicity distribution will
likely be bimodal, with a population of metal-poor clusters consisting of
the GCs formerly belonging to the halos of the now-merged disk galaxies and
a population of metal-rich clusters both stemming from these disks and
formed during the merger itself.

If so, \n3921 may be related to elliptical galaxies like \n1316
\citep{goud01a}, \n3610 \citep{whit97,stra03,stra04}, and \n5128
\citep{peng04}, which all appear to be descendants of more ancient
mergers involving disk galaxies.  Their common trait is the presence
of both fine structure indicative of past mergers and some metal-rich
globular clusters of intermediate age ($\sim$2\,--\,8 Gyr).  Though not
showing any {\it obvious} signs of tidal perturbation, the Virgo Cluster
elliptical \n4365 may belong to this category of merger descendants as
well. It, too, features metal-rich globular clusters of intermediate
age \citep{lars03} and a kinematically decoupled, slightly bluish
central disk that may stem from a past merger \citep{surm95,caro97}.

\subsection{Nature and Origin of Clusters}

What is the nature of clusters S1 and S2, and how did they form?

The spectroscopic observations presented above support the view that many
of the bluish point sources observed in \n3921 (S96; SMWF96)
indeed are young globular clusters.  In essence, any dense cluster with
$\ga$10$^4$ stars, an effective (= half-light) radius \reff\
of the order of 10~pc or less, and an age exceeding 10\,--\,20 core-crossing
times \tcross\ has to be gravitationally bound and is a globular cluster.

Assuming that S1 and S2 have \citet{chab03} IMFs, their luminosities and
ages indicate present-day masses of $(2.3\pm 0.1)\times 10^6 \msun$ and
$(1.5\pm 0.1)\times 10^6 \msun$, respectively (BC03), whence they must
contain well over 10$^6$ stars each. 
Both clusters have measured effective
radii of $\reff\la 5$~pc, implying that---like Milky-Way globulars---they
have core-crossing times of a few Myr \citep{meyl97}.  Hence, each
is of order $\sim$100\,\tcross\ old and clearly a gravitationally
bound {\it young globular cluster}.

According to the BC03 models, the two clusters have so far lost about
35\% of their original mass due to stellar evolution and will loose another
$\sim$15\% over the coming 13 Gyr due to the same cause.  At an age of
13.5 Gyr and if stellar evolution were the only source of mass loss,
S1 and S2 would be 45\% and 28\% as massive as $\omega$~Cen is
at present ($\mass_{\omega {\rm Cen}}\approx 4.0\times 10^6 \msun$,
\citealt{meyl02}). Hence, these are both dense and massive GCs by
any measure.  Although their dynamical evolution and evaporative mass loss
will depend on their unknown orbital characteristics, it seems likely
that both clusters will survive for a long time, of the order of a Hubble
time or longer.

How did S1, S2, and the many other candidate young GCs in \n3921 form?
There is growing evidence that massive young clusters in present-day
mergers form from Giant Molecular Clouds (GMC; e.g., \citealt{schw03};
\citealt{whit03}). As \citet{jog92} first pointed out and \citet{elme97}
reasoned in more detail, the rapidly mounting pressure of interstellar
gas heated in starbursts provides a natural mechanism for triggering the
collapse of the embedded GMCs, which are cold and only marginally stable.
In support of this hypothesis, the luminosity functions of young
globular-cluster systems are remarkably similar to the mass functions of
GMCs in present-day spiral galaxies \citep{harr94}.  Both kinds of
functions are power laws with similar exponents and upper-end cutoffs.
Specifically, the luminosity function of young clusters in \n3921 has
the form $\phi(L) dL \propto L^{\alpha} dL$, where $\alpha = -2.1\pm 0.3$
(SMWF96), in accord with the luminosity functions of many other
systems of young clusters and GMCs.

A further piece of evidence in favor of the young globulars of \n3921 having
formed from GMCs in the merging disk galaxies is provided by the clusters'
radial distribution.  This distribution closely follows the $r^{1/4}$-law
of \n3921's light distribution.  The agreement indicates that the progenitors 
of the young GCs experienced the same violent relaxation as did the disk
stars, which in turn suggests that these progenitors were compact and
reacted to the merger like point masses (SMWF96). 
The GMCs of the input galaxies are the only known gas-rich objects to be
both compact and massive enough to fit the bill.  Hence, the radial cluster
distribution in \n3921 strongly points to GMCs having been the progenitors
of the young globulars.  The fact that S1 and S2 have approximately solar
metallicities (\S 3.3) is consistent with this hypothesis.

\section{SUMMARY}

We have described exploratory Keck-LRIS spectroscopy of two candidate
globular clusters of apparent magnitude $V\approx 22.2$ in the merger
remnant \n3921.  The main results are as follows:

(1) Both objects are confirmed to be young star clusters of visual
absolute magnitude $M_V\approx -12.5$ within \n3921. They lie at
$\sim$5~kpc ($\sim$0.9\reff) projected distance from the center
and show halo kinematics.
Their line-of-sight velocities relative to the nucleus are $+114\pm
34$ \kms\ and $+51\pm 25$ \kms, respectively, while that of the local
galaxy background is $-25\pm 26$ \kms.

(2) The clusters' spectra feature strong Balmer absorption lines up to
H13, indicating that A-type stars dominate the main-sequence turnoffs.
Comparisons with models by \citet{bc96,bc03} and \citet{bres96}
yield ages of $450^{+80}_{-110}$ Myr for Cluster S1 and 
$280^{+70}_{-80}$ Myr for S2, and a mean metallicity that is roughly solar
(to within $\pm 0.5$~dex).  As judged by their half-light radii of
$\la$5~pc and their ages corresponding to $\sim$10$^2$ core-crossing times,
both clusters are gravitationally bound and, hence, young globulars.

(3) These two young globulars appear to be massive and dense.  Estimates
based on their luminosities and BC03 models with Chabrier (2003) initial
mass functions yield cluster masses of $(2.3\pm 0.1)\times 10^6 \msun$
and $(1.5\pm 0.1)\times 10^6 \msun$ for S1 and S2, respectively, or
roughly half the mass of $\omega$ Cen.  With such masses and normal
effective radii, both clusters are likely to survive for a period of
the order of a Hubble time or longer.

(4) If most of the newly-formed GCs in \n3921 have metallicities
comparable to those of S1 and S2, then the recent merger of two disk
galaxies has not only formed a protoelliptical, but also enhanced the
metal-rich component of the cluster population.
This enhancement is significant, since the ratio of young to old GCs
in \n3921 is $\ga$0.4 (SMWF96), while the average ratio of
metal-rich to metal-poor GCs in normal elliptical galaxies is
$0.8\pm 0.4$ ($1\sigma$ range).  Therefore, \n3921 and its GC system
provide evidence in support of the hypothesis that giant ellipticals,
and especially those with bimodal GC systems, can form through major
mergers involving gas-rich disk galaxies.

We conclude by noting that integrated-light spectroscopy of {\it old\,}
globular clusters in host galaxies at $cz\ga 5000$ \kms\ remains
beyond the reach of 8-m to 10-m telescopes.  It is the 2\,--\,6 mag
gain in cluster luminosity due to {\it youth} that has made the present
work in \n3921, as well as earlier cluster spectroscopy in \n7252
\citep{ss93,ss98} and \n1275 \citep{zepf95,brod98}, possible.

\acknowledgments

We thank Charles Sorenson and Gregory Wirth for expert assistance at
the telescope, and Gustavo Bruzual and St\'ephane Charlot for the
early release of their latest cluster-evolution models.
Two of us gratefully acknowledge partial support from the National
Science Foundation through grants AST-99\,00742 and AST-02\,05994
(FS), and AST-99\,00732 and AST-02\,06139 (JPB). 
All three authors recognize the important cultural role that the summit
of Mauna Kea has within the indigenous Hawaiian community, and are
grateful to have had the opportunity to conduct astronomical observations
from this mountain.


%
%
%
%

\clearpage

\begin{deluxetable}{lcccccccccccc}
\def\psn{\phs\phn}
\def\pnn{\phn\phn}
\tabletypesize{\scriptsize}
\tablewidth{0pt}
\tablenum{1}
\tablecolumns{13}
\tablecaption{Candidate Clusters Observed in NGC 3921}
\tablehead{
\colhead{}                       &
\colhead{Other}                  & 
\colhead{$\Delta\alpha_{2000}$\tablenotemark{b}} &
\colhead{$\Delta\delta_{2000}$\tablenotemark{b}} & 
\colhead{$r_{\rm proj}$\tablenotemark{c}}        &
\colhead{\reff\tablenotemark{c}} &
\colhead{$V_0$\tablenotemark{d}} &
\colhead{$(V\!-\!I)_0$\tablenotemark{d}} &
\colhead{$M_V$}                  &
\colhead{$cz_{\rm hel}$\tablenotemark{e}} &
\colhead{$\Delta v$\tablenotemark{f}} &
\colhead{Expos.}                 &
                                   \\
\colhead{\phn Name\phn}          &
\colhead{ID\tablenotemark{a}}    &
\colhead{(arcsec)}               &
\colhead{(arcsec)}               &
\colhead{(kpc)}                  &
\colhead{(pc)}                   &
\colhead{(mag)}                  &
\colhead{(mag)}                  &
\colhead{(mag)}                  &
\colhead{(km s$^{-1}$)}          & 
\colhead{(km s$^{-1}$)}          & 
\colhead{(s)}                    &
\colhead{$S/N$\tablenotemark{g}}  }
\startdata
S1\dotfill& \#27& $-$10.91& \phn$-$0.61& 4.6& $\la$5& 22.15& 0.74& $-$12.52&
	6042\,$\pm$\,30&    +114\,$\pm$\,34& 5400& 22.5 \\
S2\dotfill& \#44& \phn$-$3.43&   +12.51& 5.4& $\la$5& 22.29& 0.63& $-$12.38&
	5978\,$\pm$\,20& \phn+51\,$\pm$\,25& 5400& 22.0 \\
\enddata
\tablenotetext{a}{SMWF96.}
\tablenotetext{b}{Relative to NGC 3921 nucleus at\,\ $\alpha_{\rm J2000}$ =
	11:51:06.97,\,\ $\delta_{\rm J2000}$ = +55:04:43.0\,.}
\tablenotetext{c}{From SMWF96, converted to\,\ $H_0=70$ km s$^{-1}$ Mpc$^{-1}$.}
\tablenotetext{d}{From SMWF96, corrected for\,\ $A_V=0.023$\,.}
\tablenotetext{e}{Heliocentric radial velocity determined in this work.}
\tablenotetext{f}{$\Delta v = (cz_{\rm hel}-5926)/1.019767$; see text.}
\tablenotetext{g}{Average signal-to-noise ratio of continuum per resolution
	element ($\sim$5.5 \AA), measured in interval $\lambda\lambda =
	4500$\,--\,5900 \AA\ and corresponding to $S/N\approx 9.5$ \AA$^{-1}$.}
\end{deluxetable}

\begin{deluxetable}{lccccccccc}
\tablenum{2}
\tablecolumns{10}
\tablewidth{0pt}
\tabletypesize{\footnotesize}
\tablecaption{Line Equivalent Widths and Cluster Ages}
\tablehead{
\colhead{} &                 \colhead{H$\beta$}  & 
\colhead{H$\gamma$} &        \colhead{H$\delta$} &  
\colhead{H\,+\,H$\epsilon$}& \colhead{K} & 
\colhead{H8} &               \colhead{\hbgdav} &
\colhead{} &                 \colhead{} \\
\colhead{Object} &           \colhead{(\AA)} &
\colhead{(\AA)} &            \colhead{(\AA)} &
\colhead{(\AA)} &            \colhead{(\AA)} & 
\colhead{(\AA)} &            \colhead{(\AA)} &
\colhead{K/(H$\epsilon$\,+\,H8)}& \colhead{$\log$ Age\tablenotemark{a}} }
\startdata
S1\dotfill&  $12.9\pm1.0$& $12.9\pm1.0$& $12.5\pm1.1$& $12.4\pm1.6$&
       $2.7\pm0.9$&  $8.9\pm2.6$& $12.8\pm0.6$& $0.127\pm0.048$&
       $8.65_{-0.12}^{+0.07}$ \\
S2\dotfill&  $11.5\pm0.8$& $12.7\pm0.9$& $12.6\pm0.9$& $12.9\pm1.3$& 
       $1.9\pm0.7$&  $9.6\pm1.9$& $12.3\pm0.5$& $0.084\pm0.034$& 
       $8.45_{-0.15}^{+0.09}$ \\
Galaxy\tablenotemark{b}\dotfill&
       $\phn6.1\pm0.1$&  $\phn6.3\pm0.1$&  $\phn6.6\pm0.1$& $11.6\pm0.2$&  
       $5.6\pm0.1$&  $6.7\pm0.3$&  $6.3\pm0.1$& $0.309\pm0.008$& $\dots$ \\
\enddata
\tablenotetext{a}{Age expressed in years.}
\tablenotetext{b}{Patch of \n3921 galaxy light between clusters S1
	and S2, for comparison (see \S 2).}
\end{deluxetable}

\clearpage


\begin{figure}
  \includegraphics[angle=-90,scale=0.75]{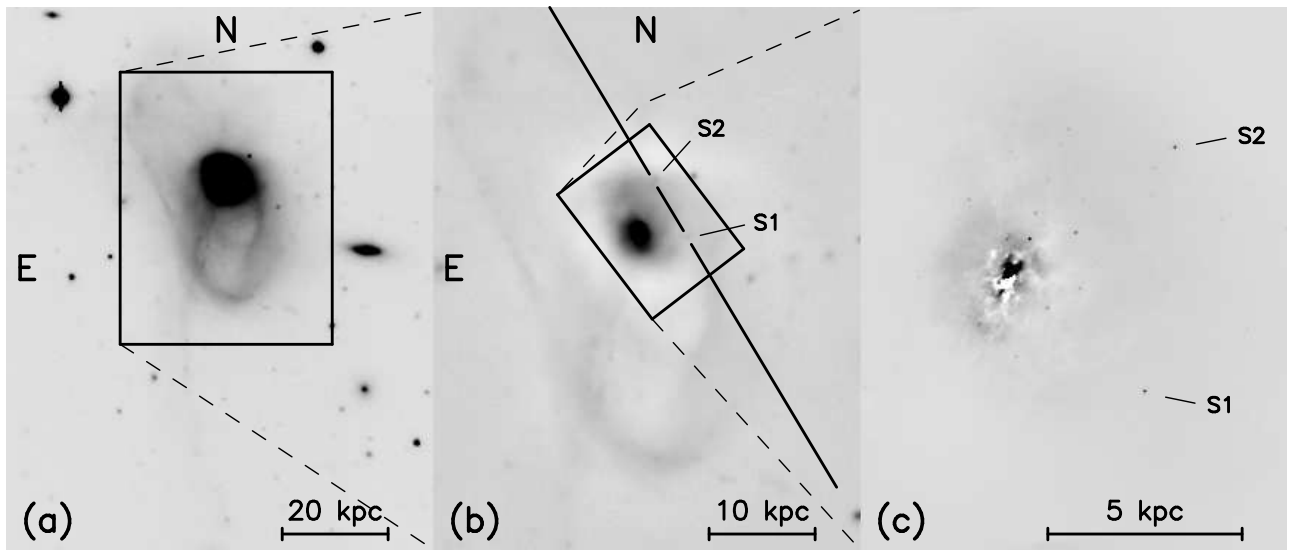}
  \caption{
Images of \n3921 with observed candidate clusters S1 and S2 marked.  The
left and middle images were obtained with the Hale 5-m telescope (S96),
and the right image with \hst/WFPC2 (SMWF96).  Boxes plus dashed lines
indicate the field of view of the next panel to the right.  The inclined
line in Panel (b) 
marks the position of the LRIS spectrograph slit during the observations.
  \label{fig1}}
\end{figure}

\clearpage

\begin{figure}
  \plotone{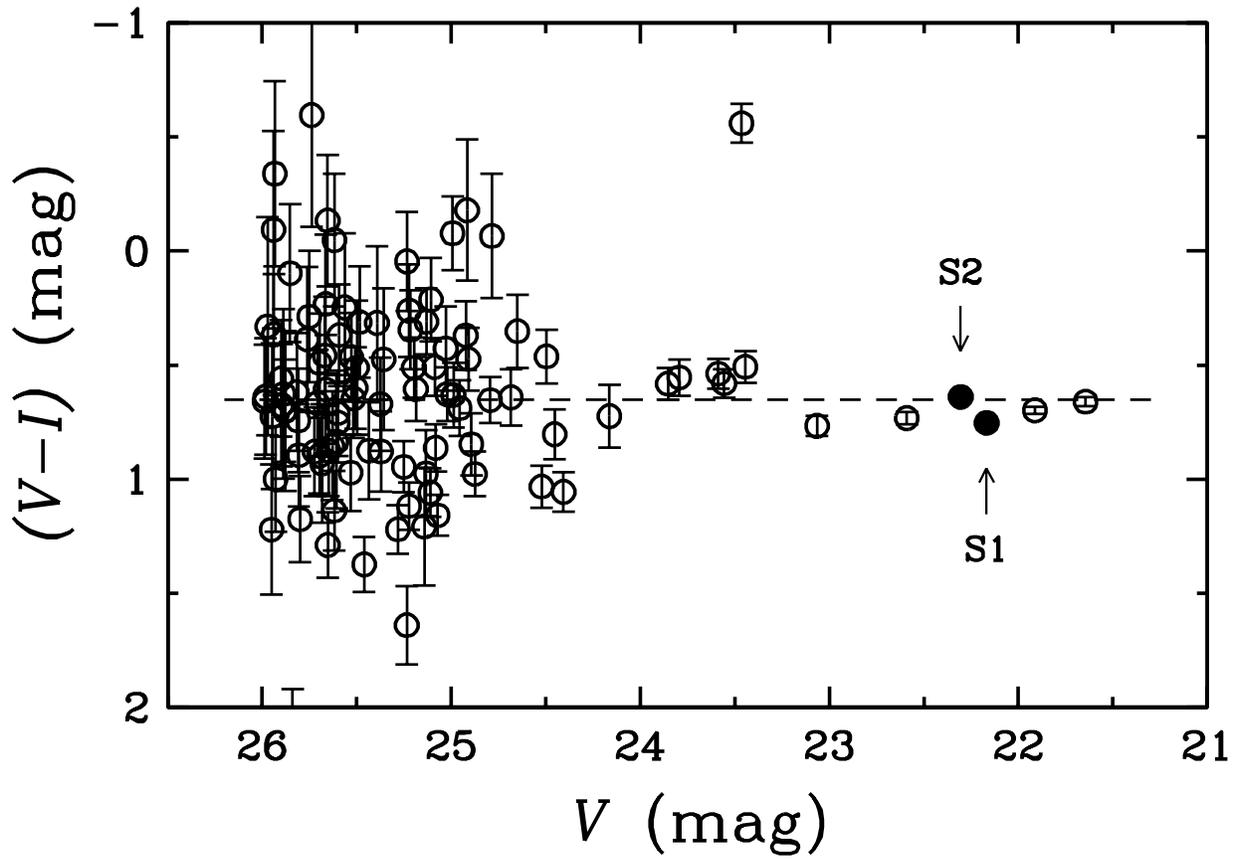}
  \caption{
Color index $(V\!-\!I)$ plotted vs magnitude $V$ for 102 candidate GCs in
\n3921 (after SMWF96).  The two candidate GCs observed spectroscopically,
S1 and S2, are marked.
  \label{fig2}}
\end{figure}

\clearpage

\begin{figure}
  \epsscale{0.8}
  \plotone{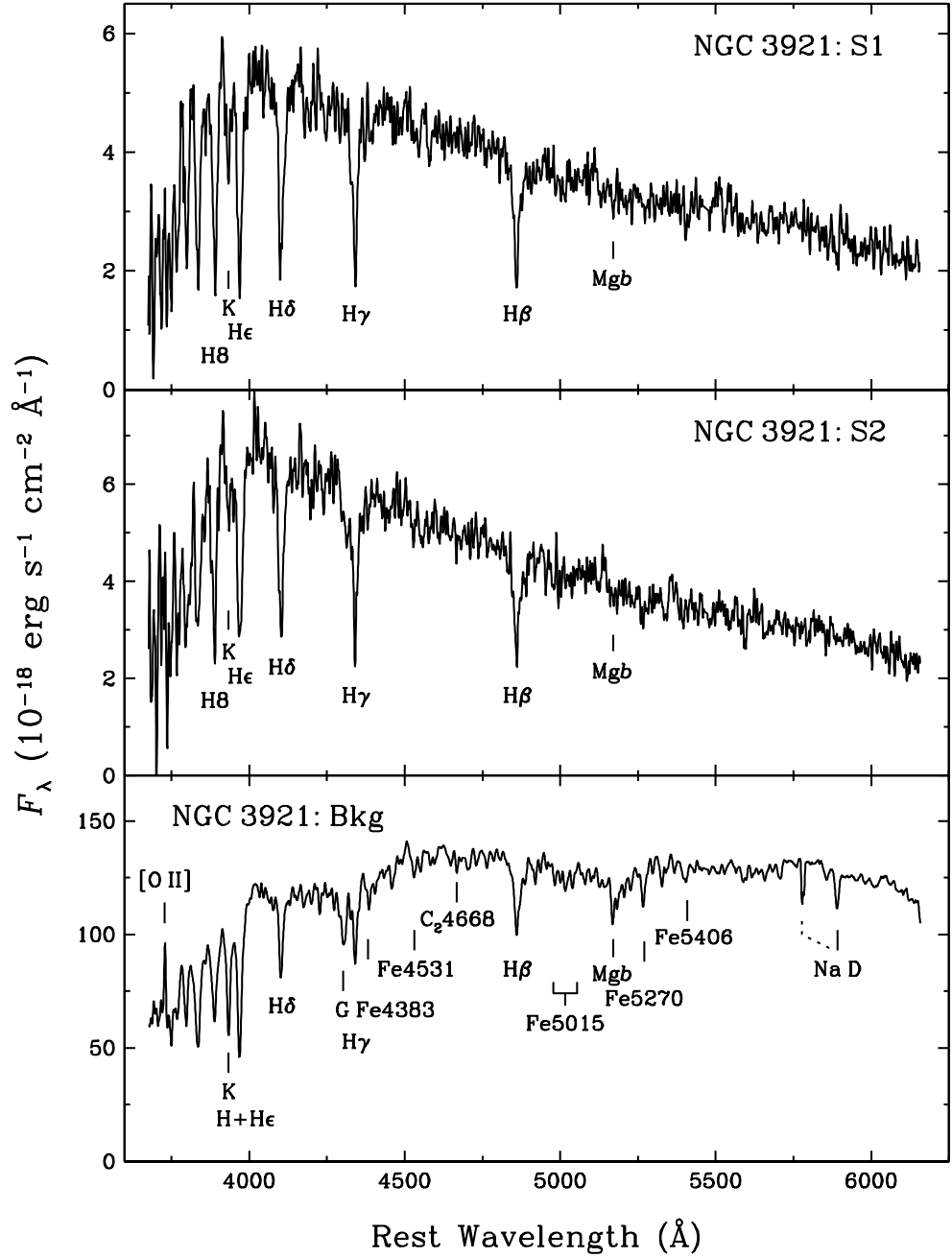}
  \caption{
Ultraviolet-to-visual spectra of Clusters NGC\,3921:\,S1, S2, and galaxy
background in between, obtained with Keck~I telescope plus LRIS through 
a $165\arcsec\times 1\farcs0$ slit.  The spectra are flux calibrated,
slightly Gaussian-smoothed ($\sigma = 1.5$ \AA) for better display, and
plotted versus rest wavelength.  Note strong Balmer absorption lines
indicative of A-type main-sequence stars in the two cluster spectra, and 
later-type, more composite nature of galaxy-background spectrum.  The
dotted line at Na\,D points to a Milky-Way foreground-absorption component.
  \label{fig3}}
\end{figure}
 
\clearpage

\begin{figure}
  \epsscale{0.8}
  \plotone{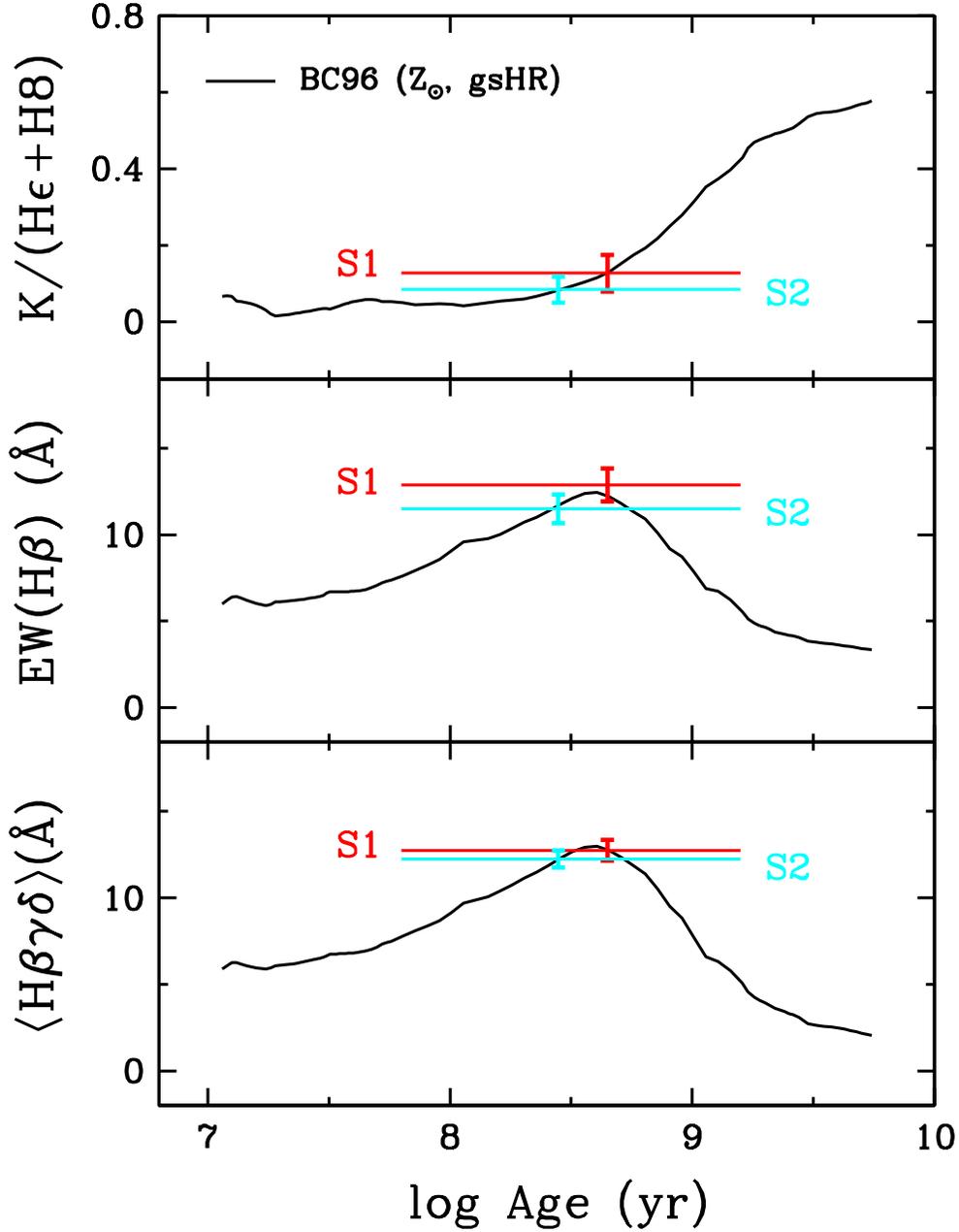}
  \caption{
Evolution of line ratio \kratio\ and equivalent widths EW(H$\beta$) and
\hbgdav\ in model-cluster spectra with age ({\it solid curves}, computed
from BC96 models of solar metallicity), compared with values measured
from spectra of clusters \n3921:\,S1 and S2 ({\it horizontal lines}).
Note that the ratio \kratio\ favors the higher of two possible
Balmer-line ages (based on \hbgdav) for S1, but the lower for S2.
  \label{fig4}}
\end{figure}

\clearpage

\begin{figure}
  \includegraphics[angle=-90,scale=0.62]{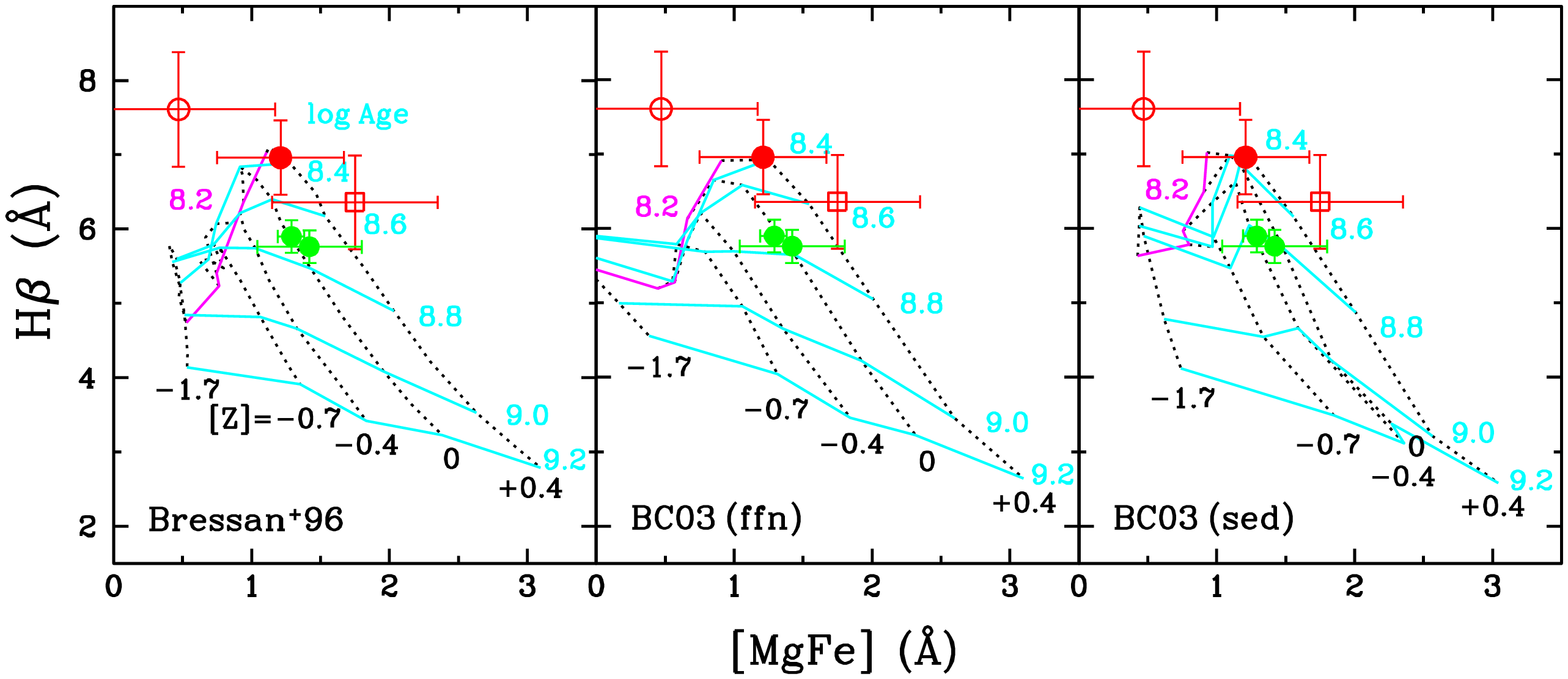}
  \caption{
H$\beta$ vs [MgFe] diagrams for clusters NGC\,3921:\,S1 ({\it red circle
with error bars}), S2 ({\it red square}), and their average ({\it red
filled circle}). Three panels show grids of isochrones ({\it blue solid
lines}) and isometallicity lines ({\it dotted}) based on models by
({\it left}) \citet{bres96},
({\it middle}) \citet{bc03} with fitting functions, and
({\it right}) \citet{bc03} with direct extraction from
spectral energy distributions.  For greater clarity, the $\log$\,age~=
8.2 isochrones are drawn in magenta.  Green data points show two young
globular clusters of \n7252 for comparison \citep{ss98}.
From this diagram, the mean age of the two \n3921 clusters appears to be
approximately 250 Myr and the mean metallicity about solar to twice solar,
both with large error bars.
  \label{fig5}}
\end{figure}

\clearpage

\begin{figure}
  \includegraphics[angle=-90,scale=0.65]{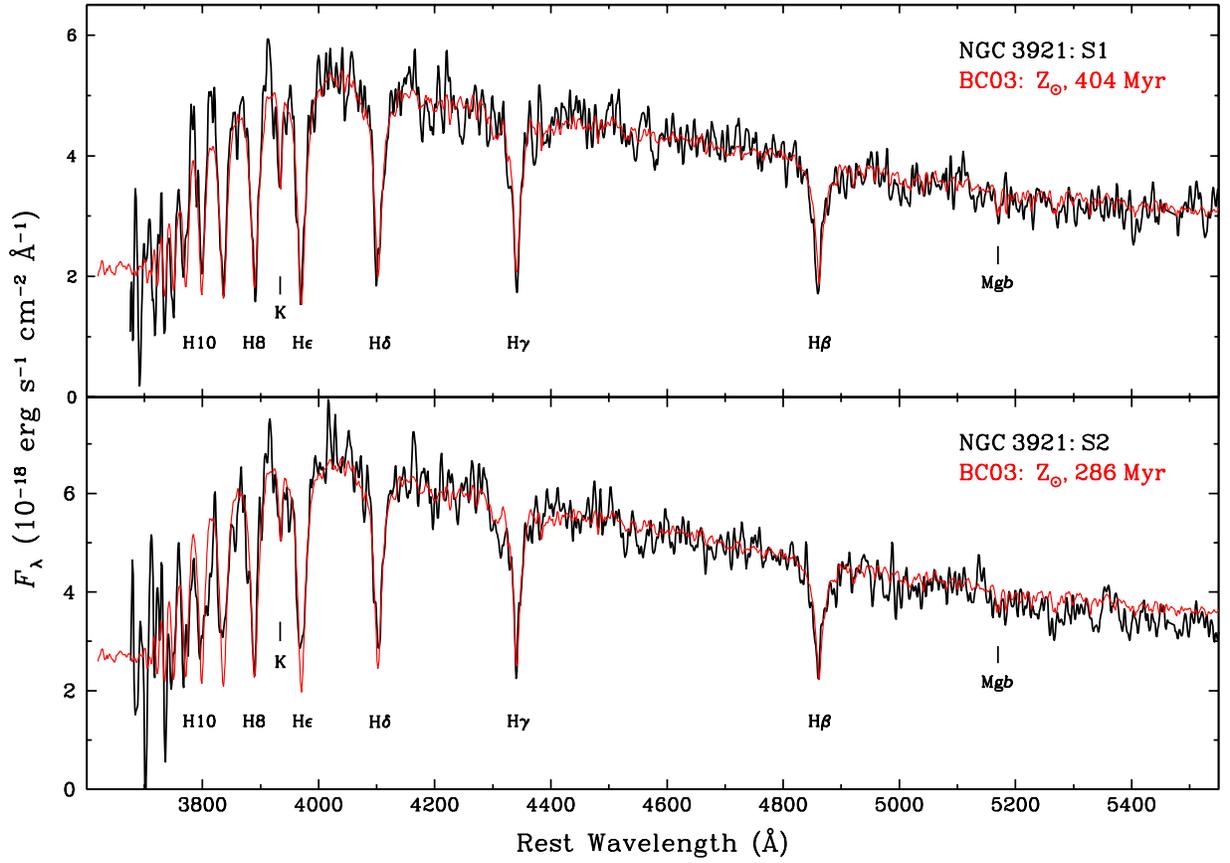}
  \caption{
Comparison between observed spectra of Clusters NGC\,3921:\,S1, S2
({\it black lines}) and model spectra for clusters of solar metallicity
({\it red lines}) by Bruzual \& Charlot (2003).  The model spectra are
unsmoothed and were chosen from the model age sequence to be closest in
time to the cluster age.  They match the more noisy observed spectra
remarkably well.
  \label{fig6}}
\end{figure}

\end{document}